\documentclass[%
 reprint,
 amsmath,amssymb,
 aps,
]{revtex4-2}

\usepackage{graphicx}
\usepackage{dcolumn}
\usepackage{bm}
\usepackage{float}

\begin{document}

\preprint{APS/123-QED}

\title{Criticality of the viscous to inertial transition near jamming in non-Brownian suspensions}

\author{Nishanth Murugan\textsuperscript{1}}
\author{Donald Koch\textsuperscript{2,1}}
\author{Sarah Hormozi\textsuperscript{2,1}}

\affiliation{\\ \textsuperscript{1}Sibley School of Mechanical and Aerospace Engineering, Cornell University, Ithaca, NY, 14853, USA \\ 
\textsuperscript{2}Robert Frederick Smith School of Chemical and Biomolecular Engineering, Cornell University, Ithaca, NY 14853, USA}

\date{\today}
\begin{abstract}
In this work, we use a Discrete Element Method (DEM) to explore the viscous to inertial shear thickening transition of dense frictionless non-Brownian suspensions close to jamming. This transition is characterized by a change in the steady state rheology of a suspension with increasing shear rate ($\dot{\gamma}$), from a regime of constant viscosity at low shear rates to a regime where the viscosity varies linearly with the shear rate. Through our numerical simulations, we show that the characteristic shear rate associated with this transition depends sensitively on the volume fraction ($\phi$) of the suspension and that it goes to zero as we approach the jamming volume fraction ($\phi_m$) for the system. By attributing the criticality of this transition to a diverging length scale of the microstructure as $\phi \rightarrow \phi_m$, we use a scaling framework to achieve a collapse of the rheological data associated with the viscous to inertial transition. A series of tests conducted on the system size dependence of the rheological results is used to show the existence of this microstructural length-scale that diverges as the suspension approaches jamming and its role in triggering the viscous to inertial transition.   
\end{abstract}
\maketitle
\begin{figure}[h]
\includegraphics[width=0.5\textwidth]{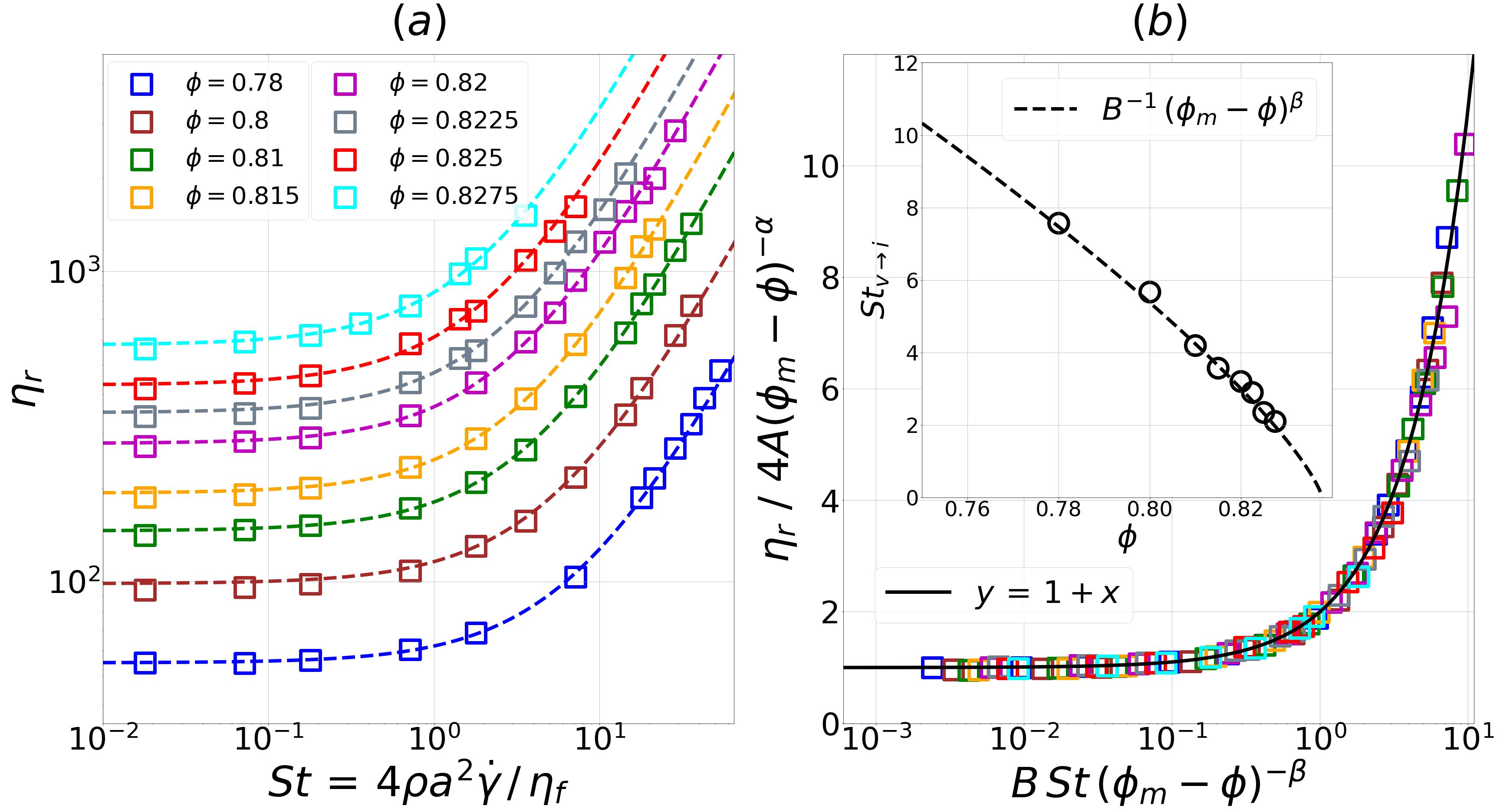}
\caption{\label{Fig:2D-R} Results from the 2D simulations. (a) Plot of the relative shear viscosity as a function of $St$ for different volume fractions $\phi$. The dashed lines depict the fitting function $\eta_r(St) \, = \, 4 \, \mathcal{A}(\phi) \, (1 \,+\, \mathcal{B}(\phi) St)$ for each $\phi$. (b) Collapse of the rheological data from the scaling framework which yields, $ \{ \, A, \, B, \, \alpha, \, \beta, \, \phi_m \, \} \,\,= \,\, \{ \, 0.283, \, 0.015, \, 1.36, \, 0.774, \, 0.838 \, \}$. The solid black line is $\sim 1 + St_{Cluster}$. In the inset, the circles correspond to $St_{v \rightarrow i}$ obtained by fitting the data for each $\phi$ in (a) separately. The solid black line in the inset depicts the scaling as inferred from the collapse of the data.}
\end{figure}
\begin{figure}[H]
\includegraphics[width=0.5\textwidth]{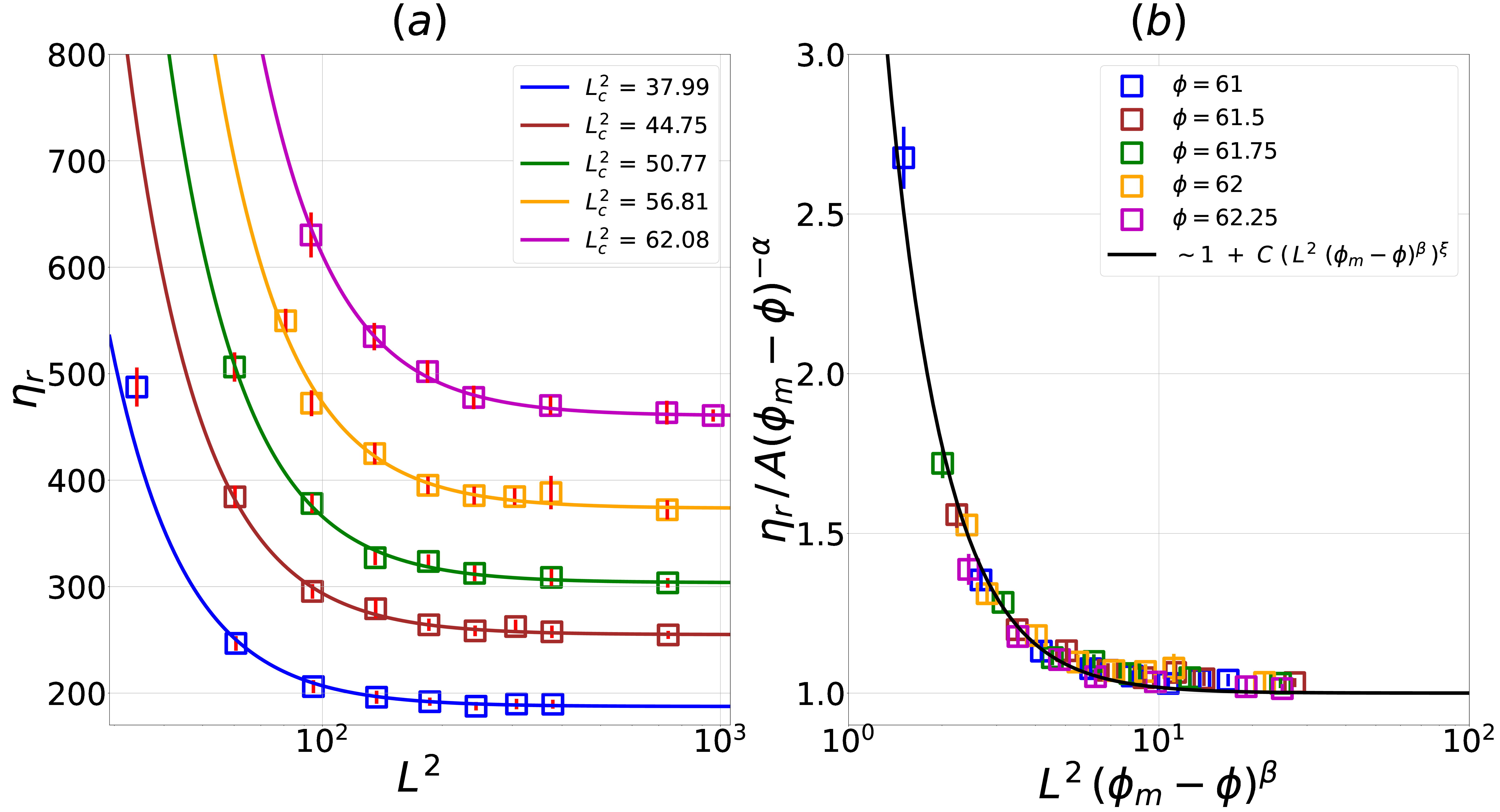}
\caption{\label{Fig:3D-N} The results for the system size dependence in the 3D system at $St=0.18$. (a) The square points depict the shear viscosity of the system as a function of the box size. The solid lines depict a power law fit for the points. (b) A collapse of the points from figure \ref{Fig:3D-N}(a) is achieved using the results on the scaling obtained from the rheology. The solid black line represents a power law fit for the collapsed points with, $\xi = 2.51$.}
\end{figure}

The rheology of dense non-Brownian suspensions is central to the physics of various natural phenomena (mudslides, landslides, avalanches, etc) and is of critical importance in the industrial processing of cement slurry, ceramic materials, etc. With increasing shear rate the suspension stress displays a transition from a viscous regime of Newtonian rheology ($\propto \dot{\gamma}$) to an inertial regime characterized by Bagnoldian rheology ($\propto \dot{\gamma}^2$). This behavior is attributed to arise from the increasing role played by particle inertia in facilitating the transfer of momentum within the suspension. However, the characteristic shear rate for the transition is found to vary across different systems in the literature and across different works without any consensus on the physical mechanism that triggers the transition. 
\\
In the viscous regime of constant viscosity at low shear rates, there is very minimal influence of particle inertia on the dynamics of the system, with the particle contact forces being balanced by the hydrodynamic lubrication forces within the suspension. As a result, the stress in this regime scales linearly with the shear rate and is given by, $\Sigma_V \, \sim \, \eta_V(\phi) \, \eta_f \, \dot{\gamma}$, where $\eta_f$ is the viscosity of the solvent phase. In the inertial regime wherein inter-particle contact or repulsive forces between particles control the transfer of momentum, the scaling for the stress is written as $\Sigma_I \, \sim \, \eta_I(\phi) \,  \rho \, a^2  \eta_f  \dot{\gamma}^2 $. Intuitively, one would expect the transition to occur when the viscous and inertial scaling for the stresses within the suspension are similar in magnitude. This results in the definition of a non-dimensional Stokes number given by  $St = 4 \rho a^2 \dot{\gamma} \, / \, \eta_f$, which is a measure of the relative strength of inertial and viscous stresses within the suspension at the particle scale. Here, $\rho$ and $a$ represent the density and average radius of a particle in the suspension. This leads one to predict the transition to occur at a characteristic Stokes number, $St_{v \rightarrow i} \approx 1$. However, experimental \cite{fall2010shear,madraki2020shear,tapia2022viscous,le2023solvents} and numerical \cite{trulsson2012transition,ness2016shear,vo2020additive} investigations into the viscous to inertial shear thickening transition in dense suspensions have yielded varied results on $St_{v \rightarrow i}$. 
\\
In the works of \cite{fall2010shear,madraki2020shear,le2023solvents}, the experiments are performed on a suspension of polystyrene spheres that are roughly $\sim 40 \, - \, 80 \, \mu m$ in size immersed in an aqueous solution. The atomic force microscopy (AFM) measurements on samples of this system \cite{madraki2020shear,le2023solvents} confirm the presence of a strong electrostatic repulsive force between the particles which in turn renders the suspension to be nearly frictionless. The experiments \cite{fall2010shear,madraki2020shear}, in addition to the asymptotic scaling theories close to jamming for frictionless systems \cite{degiuli2015unified}, show that $St_{v \rightarrow i}$ varies as a function of $\phi$ and that it goes to zero as the suspension approaches jamming. 
\\
In contrast, the work of \cite{tapia2022viscous} uses a suspension made up of PMMA particles that are roughly $\sim 4.65 \, mm$ in size. This yields a system that is frictional due to the length scale associated with the colloidal forces being much smaller than the roughness of the particles, resulting in particles coming into physical contact during the flow of the suspension. The rheology of this system yields $St_{v \rightarrow i} \approx 10$, a result that is independent of the volume fraction of the suspension. Furthermore, numerical simulations on the viscous to inertial transition in frictional dense suspensions have also shown the transition to be independent of $\phi$, $St_{v \rightarrow i} \approx 1$  \cite{ness2016shear} and $St_{v \rightarrow i} \approx 0.01 $  \cite{vo2020additive}. These results highlight the stark differences between the microstructure of frictional and frictionless suspensions. More interestingly, the variation in the Stokes number of transition with volume fraction in the case of frictionless suspensions hints towards the possible existence of a length scale other than the particle size that controls the transition. 
\\ 
In this letter, we use a Discrete Element Method to examine the rheology of a dense non-Brownian frictionless suspension over a wide range of shear rates and volume fractions. We show that the characteristic Stokes of transition $St_{v \rightarrow i}$ in a frictionless suspension goes to zero as a nearly linear function of the distance to the jamming volume fraction $(\phi_m - \phi)$. We attribute the changes in $St_{v \rightarrow i}$ with $\phi$ to arise from the existence of a growing length scale of the microstructure. We prove this by showing that the diverging length scale of the microstructure and the critical behavior of the transitional Stokes number obey the same scaling as a function of the distance to jamming $(\phi_m - \phi)$.
\begin{figure}[h]
\includegraphics[width=0.5\textwidth]{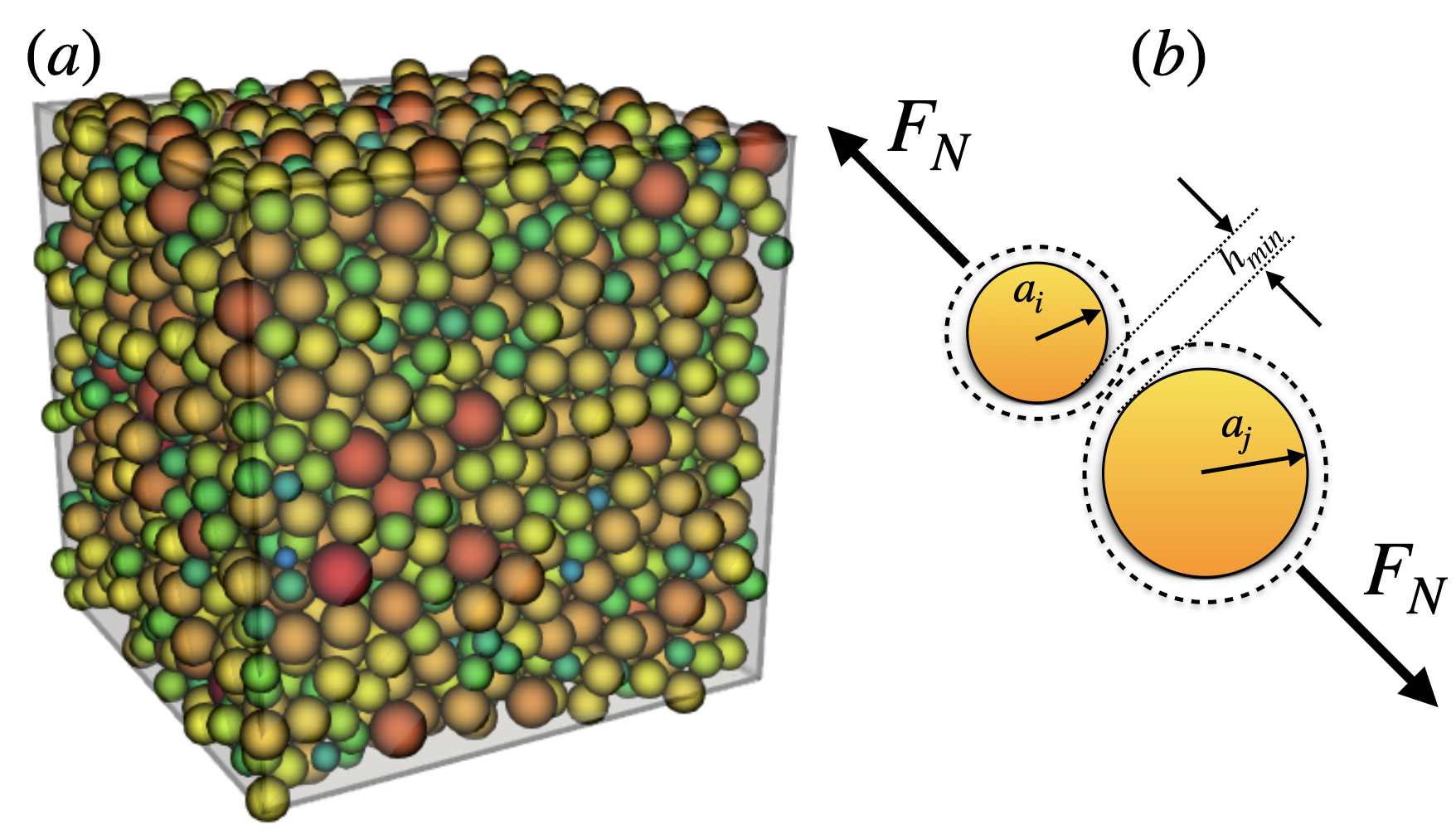}
\caption{\label{fig:schematic} (a) A snapshot of the DEM simulation for a polydisperse suspension of particles wherein the color gradient is indicative of the particle size. (b) A schematic of the electrostatic repulsive barrier between particles in frictionless suspensions}
\end{figure}
\\
\textit{Numerical Method:} The discrete element method solves Newton's equations of motion (\ref{Eqn:DEM}) for each particle within the suspension by computing its interaction with all other particles in the system. In frictionless non-Brownian suspension, two types of forces majorly dictate the dynamics of the suspension:  1) a hydrodynamic lubrication force $\bm{\mathcal{F}}^L_{ij}$ caused by the flow within the gap between two particles due to their relative motion, and 2) a repulsive or contact force $\bm{\mathcal{F}}^R_{ij}$ caused by either an electrostatic interaction or a Hertzian elastic contact between two particles. In a frictionless system, the interaction torque between two particles comes purely from the lubrication flow within the gap between two particles ($\bm{\mathcal{T}}^L_{ij}$). The pair-wise nature of these two interactions thus simplifies the DEM in needing to only account for interactions between a particle and its neighbors. The equations of motion are,
\begin{gather} \label{Eqn:DEM}
    \begin{gathered}
    m \frac{d \bm{u}_i}{dt} \,\, = \,\, \sum_{j=1}^{N_{n,i}}  \Big( \bm{\mathcal{F}}^L_{ij} \,\,+\,\,  \bm{\mathcal{F}}^R_{ij} \Big), 
    \hspace{0.2cm}
    I \frac{d \bm{\omega}_i}{dt} \,\, = \,\, \sum_{j=1}^{N_{n,i}}   \bm{\mathcal{T}}^L_{ij},
    \\
    \frac{d\bm{x}_i}{dt} \,\, = \,\, \bm{u}_i. 
\end{gathered}
\end{gather}
In the above equation, $\bm{x}_i$, $\bm{u}_i$ and $\bm{\omega}_i$ represent the position, velocity and angular velocity of the $i^{\text{th}}$ particle within the suspension. The net force and torque on the $i^{\text{th}}$ particle in the system is calculated by summing its force and torque contributions arising from interactions with all $N_{n,i}$ neighbors. In the near frictionless suspensions used in experiments \cite{fall2010shear,madraki2020shear,le2023solvents}, there is an electrostatic repulsive force that works to prevent particles within the suspension from coming into contact resulting in a finite surface separation $h$. Owing to the expensive computational requirements associated with resolving such a small Debye length scale $(\lambda \,\approx \, 2 \,  nm$ as found in \cite{le2023solvents}) for the electrostatic repulsion in a simulation with a large system size, the repulsive force between two particles is modeled as a Hookean spring, $\bm{\mathcal{F}}^R_{ij} \, = \, \mathcal{K} \, \delta \, \bm{n}$. Here, $\mathcal{K}$ is the stiffness of the spring, $\bm{n}$ is a unit vector along the line connecting the centers of two contacting particles, and, $\delta$ measures the strength of the repulsive force based on the surface separation,
\begin{gather}
    \delta =
  \begin{cases}
    0 & h>h_{min},\\
    h_{min} - h  & h<h_{min}.\\
  \end{cases}
\end{gather}
The Hookean spring activates when $h < h_{min}$. This is representative of the electrostatic repulsive force between the particles in the experiments wherein there is a finite surface separation $h$ at which the normal motion between the particles is constrained by the repulsive barrier. The expressions for the lubrication forces $\bm{\mathcal{F}}^L_{ij}$ and torques $\bm{\mathcal{T}}^L_{ij}$, that contain contributions from the squeeze, shear, and pump modes can be found in the supplementary material. However, these asymptotic expressions diverge when the surface separation between the particles ($h$) goes to zero as $\sim 1/h $ for the squeeze mode and $\sim \log(1/h)$ for the shear and pump modes. In the frictionless suspension, the presence of an electrostatic repulsive barrier constrains the normal motion between two particles by preventing them from coming closer than $h_{min}$. However, the approximation of the repulsive force as a Hookean spring in the simulations allows for two particles to come closer than $h_{min}$. In the simulations, the lubrication force is thus calculated assuming that the surface separation is $h_{min}$ when $h < h_{min}$. 
\\
The dynamics of the system governed by equation \ref{Eqn:DEM} is thus dictated by four parameters: the volume fraction of the suspension $\phi$, the Stokes number $St\,=\, 4\rho a^2 \dot{\gamma} \, / \, \eta_f $, the surface separation at which the electrostatic repulsive forces constrain normal motion between the particles given by $h_{min}$ and a non-dimensional number $\Gamma = \mathcal{K} \, \, / \, \, 6 \pi \eta_f a \,\dot{\gamma}$ \cite{gallier2014rheology} that characterizes the strength of the hydrodynamic lubrication forces, with respect to the repulsive forces. For the frictionless suspension of non-Brownian spheres, $40 \, \mu m$ in diameter,  as used in \cite{fall2010shear,madraki2020shear,le2023solvents}, the Debye length of the system measured using AFM, $\lambda \approx 10^{-4} a$ \cite{le2023solvents}. In this work, we fix the value of $h_{min} = 10^{-4}$. It is essential to have $\Gamma \gg 1$ to impose a strong resistance to normal motion between particles when the surface separation falls below $h_{min}$. In the simulations, we choose values of $\Gamma$ to vary from $10^5 - 10^7$ (\cite{goyal2024flow}) ensuring that we are firmly within the regime wherein increasing the stiffness of the Hookean spring further has a negligible effect on the rheology of the suspension. The details on the role of $\Gamma$ on the rheology of the suspension have been relegated to the supplementary material. This allows us to explore the rheology of the system within the space of $(\phi,St)$.
\\
Within the framework of the discrete element method described thus far, we solve the equations of motion (\ref{Eqn:DEM}) for each particle within the suspension, with the simulation domain being subjected to a Lees Edwards boundary condition, which is representative of a periodic domain for a shear flow. The particle polydispersity of the system is fixed at $16 \%$ to avoid the effects of crystallization. The simulations are performed in both 2D and 3D. In this work, the 2D simulations are run by constraining all particles in the system to have zero force along the vorticity direction of the shear flow. Setting the velocity along the vorticity direction and angular velocities in the flow-gradient plane to zero at the start of the simulation yields a two-dimensional shear flow of the suspension. 
\\ \textit{Results:}
The simulation results for the macroscopic rheology of the 2D system are displayed in figure \ref{Fig:2D-R}(a). The curves for different values of $\phi$ demonstrate a regime of constant viscosity at low $St$, followed by a transition to a regime where the viscosity varies linearly with $St$. The dashed lines represent the fitting function, $\eta_r(\phi,St) \, = \, 4 \, \mathcal{A}(\phi) \, (1 \,+\, \mathcal{B}(\phi) St)$. The characteristic Stokes of transition given by $St_{v \rightarrow i}(\phi) \, = \, 1 \, / \, \mathcal{B}(\phi) $, is marked by the viscous and inertial stresses being equal in magnitude. The variation of $St_{v \rightarrow i}$ as a function of $\phi$ (see inset in figure \ref{Fig:2D-R}(b)) indicates that inertia on the scale of the particle size is not the determining factor in triggering the transition to an inertial regime. It also hints at the existence of a microstructural length scale much larger than the size of a single particle, with the role played by inertial effects on the scale of such a structure being a plausible cause for the transition to the inertial regime. The growth in the length scale of such a structure with increasing $\phi$, can be used to rationalize the corresponding fall in the characteristic shear rate of the transition or $St_{v \rightarrow i}$. 
\begin{figure}[h]
\includegraphics[width=0.5\textwidth]{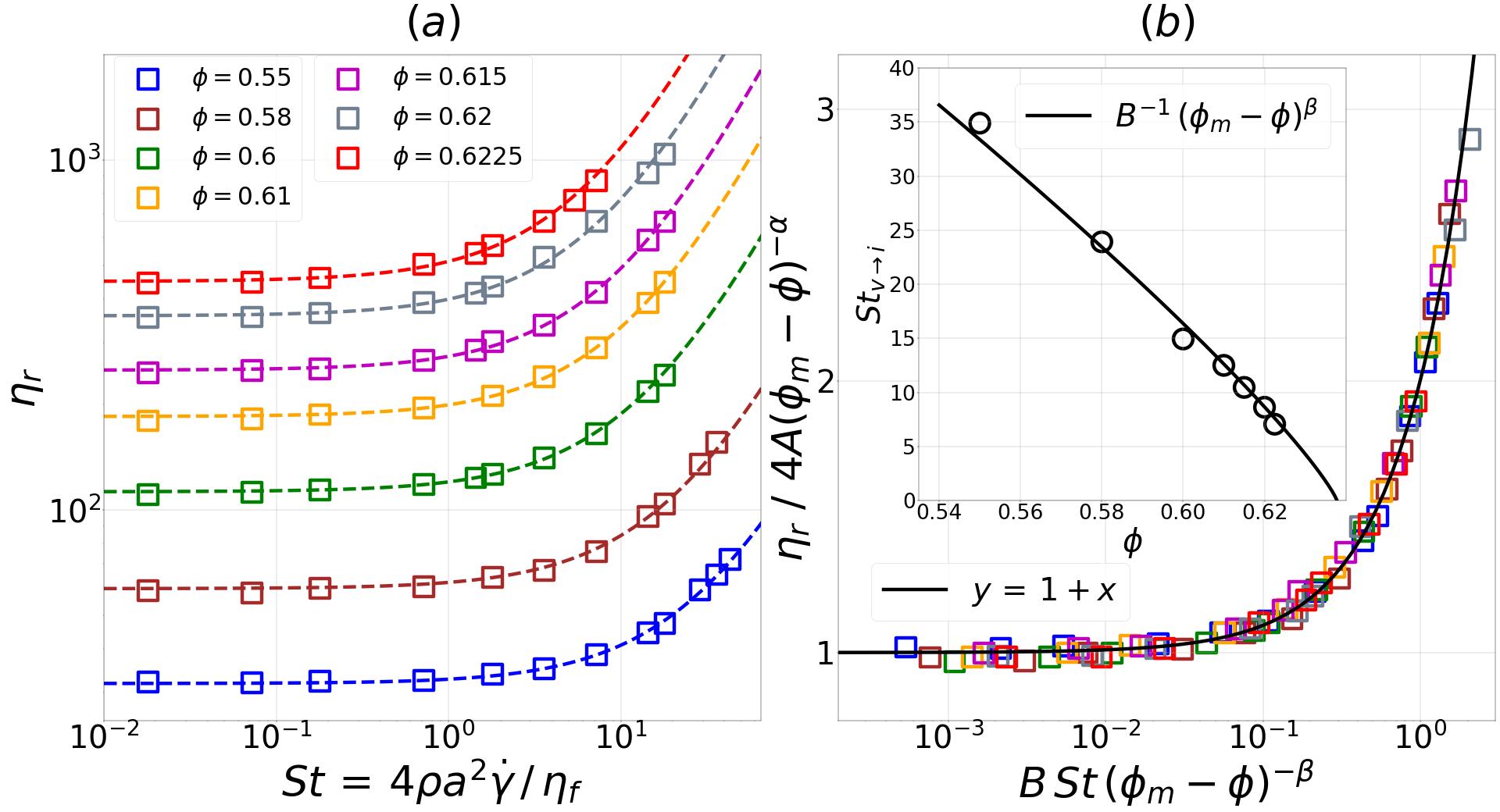}
\caption{\label{Fig:3D-R}Results from the 3D simulations. (a) Plot of the relative shear viscosity as a function of $St$ for different volume fractions $\phi$. The dashed lines depict the fitting function $\eta_r(St) \, = \, 4 \, \mathcal{A}(\phi) \, (1 \,+\, \mathcal{B}(\phi) St)$ for each $\phi$. (b) Collapse of the rheological data from the scaling framework which yields, $ \{ \, A, \, B, \, \alpha, \, \beta, \, \phi_m \, \} \,\,= \,\, \{ \, 0.193, \, 0.004, \, 1.53, \, 0.845, \, 0.638 \, \}$. The solid black line is $\sim 1 + St_{Cluster}$. In the inset, the circles correspond to $St_{v \rightarrow i}$ obtained by fitting the data for each $\phi$ in (a) separately. The solid black line in the inset depicts the scaling as inferred from the collapse of the data.}
\end{figure}
By considering the phenomenon of jamming to be accompanied by the presence of a diverging length scale within the microstructure \cite{olsson2007critical,ness2022physics}, it is possible to put forth the following scaling framework for the shear viscosity of the suspension,
\begin{gather}
    \eta_r(\phi,St) \, =  \, 4 A  (\phi_m - \phi)^{-\alpha} \big( 1  +  B \, St \, (\phi_m-\phi)^{-\beta}\big).
\end{gather}
This framework incorporates a scaling for the shear viscosity given by a constant $\eta_r \sim (\phi_m - \phi)^{-\alpha}$ at low shear rates in the viscous regime, and a scaling of $\eta_r \sim (\phi_m - \phi)^{-(\alpha+\beta)} \, St$ in the inertial regime at large shear rates. It also allows for the definition of a modified Stokes number, $St_{Cluster} = \rho_p L_c^2 \dot{\gamma} / \eta_f = B \, St \, (\phi_m-\phi)^{-\beta}$, defined based on a length scale of a cluster within the microstructure that grows and diverges as $L_c \sim d\, (\phi_m - \phi)^{-\beta/2}$. This procedure allows for a collapse of the rheological data from figure \ref{Fig:2D-R}(a) as shown in figure \ref{Fig:2D-R}(b). A similar set of results is derived for the 3D simulations as shown in figure \ref{Fig:3D-R}. The scaling shows the critical behavior of the transition to inertial shear thickening, wherein the characteristic shear rate of transition goes to zero as $\phi \rightarrow \phi_m$ with an exponent $\beta = 0.79$ and $0.84$ in 2D and 3D respectively. These results indicate that a suspension at jamming is always inertial. From the values of $\beta$, we can predict the microstructural length scale to diverge at jamming with an exponent close to $\sim -\,0.4$. However, while the scaling collapse showcases the criticality associated with the viscous to inertial transition as the system approaches jamming, it does not prove the existence of a microstructural length scale that governs the transition. \\
\begin{figure}[h]
\includegraphics[width=0.5\textwidth]{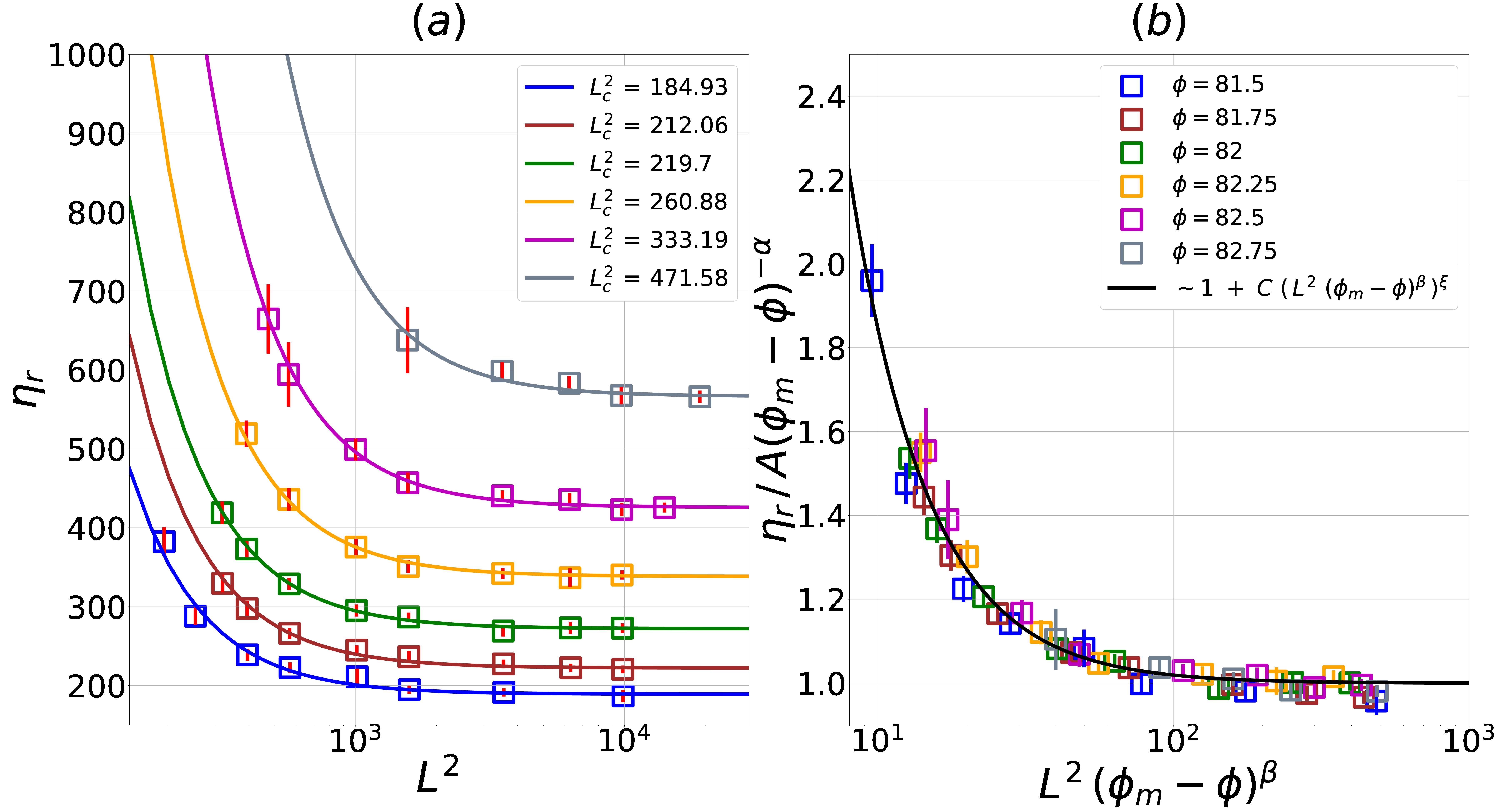}
\caption{\label{Fig:2D-N} The results for the system size dependence in the 2D system at $St=0.018$. (a) The square points depict the shear viscosity of the system as a function of the box size. The solid lines depict a power law fit for the points. (b) A collapse of the points from figure \ref{Fig:2D-N}(a) is achieved using the results on the scaling obtained from the rheology. The solid black line represents a power law fit for the collapsed points with, $\xi = 1.58$.}
\end{figure}
 To extract an approximate length scale of the cluster within the suspension, a series of tests are performed to probe the system-size dependence of the rheology at different values of $\phi$. The results of these tests for the 2D system are shown in figure \ref{Fig:2D-N}(a). When the system size is smaller than the size of the cluster, there is a collision of the cluster with itself via reaching through the periodic boundary of the box. This results in an increasing viscosity with decreasing system size of the simulation as seen in figure \ref{Fig:2D-N}(a). The results from the system size dependence can be fit by a power law, $\eta_r(\phi, L^2) = \mathcal{R}(\phi) \big(1+(L^2 \, / \, L_c^2)^{\xi}\big)$ to extract an approximate length scale for the cluster $(L_c)$ at a given $\phi$ and box size $L$. 
From the results on the rheology given in figures \ref{Fig:2D-R} and \ref{Fig:3D-R}, we expect the size of the cluster $L_c$ to scale as $L_c^2 \sim (\phi_m - \phi)^{-\beta}$ with the appropriate $\beta$ from the 2D and 3D simulations. In addition, $\mathcal{R}(\phi)$ which depicts the system size independent rheology is already known from figures \ref{Fig:2D-R} and \ref{Fig:3D-R}. For the results in figure \ref{Fig:2D-N}(a) which corresponds to the viscous regime, the power law simplifies as, 
\begin{gather}
    \eta_r(\phi, L^2) = A(\phi_m - \phi)^{-\alpha} \Big[1+ C \big(L^2 (\phi_m - \phi)^{\beta}\big)^{\xi}\Big].
\end{gather}
Scaling $\eta_r$ and $L^2$ using the values of $A, \, \phi_m, \, \alpha$ and $\beta$ known from previous results, it is possible to collapse the data in figure \ref{Fig:2D-N}(a) into a single curve as shown in figure \ref{Fig:2D-N}(b). A similar exercise can be repeated for the 3D results as shown in figure \ref{Fig:3D-N}. The collapse of the results on the system size dependence on the rheology, using the scaling for the cluster size determined from the results on the rheology of transition to inertial shear thickening proves the existence of a cluster that governs the viscous to inertial transition. 
In summary, using DEM simulations we have shown that the transition from a viscous regime to an inertial regime in a dense non-Brownian frictionless suspension displays a critical behavior as the suspension approaches jamming,  $St_{v \rightarrow i} \sim (\phi_m - \phi)^{-\beta}$. The values for the exponent $\beta = 0.79 \,\, \& \,\, 0.85$ as obtained from the 2D and 3D simulations respectively align well with the experimentally obtained results of $St_{v \rightarrow i} \sim (\phi_m - \phi)^{-1}$ \cite{fall2010shear}. It is further shown that the criticality of $St_{v \rightarrow i}$ arises from a growing length scale of a cluster within the suspension that diverges at jamming, $L_c \sim (\phi_m - \phi)^{-\beta/2}$. Such a structure within the suspension strongly highlights the non-local rheology of the system. This strongly motivates the need for a non-local theory that accounts for the transmission of stress through the suspension during the formation and destruction of the cluster, and its interactions with other particles within the suspension.
\\
However, it is necessary to note that the presence of frictional contacts within the suspension builds a microstructure that takes away the criticality of the transition seen in frictionless suspensions \cite{tapia2022viscous}. Future studies on the scaling for the cluster size close to jamming in frictional suspensions and its role in triggering the transition are essential for establishing a unified theory across different flowing regimes.
\\
This work was supported by the NSF grant CBET-1554044. The authors report no conflict of interest.
\bibliography{Ref}

\providecommand{\noopsort}[1]{}\providecommand{\singleletter}[1]{#1}%
\begin{thebibliography}{12}%
\makeatletter
\providecommand \@ifxundefined [1]{%
 \@ifx{#1\undefined}
}%
\providecommand \@ifnum [1]{%
 \ifnum #1\expandafter \@firstoftwo
 \else \expandafter \@secondoftwo
 \fi
}%
\providecommand \@ifx [1]{%
 \ifx #1\expandafter \@firstoftwo
 \else \expandafter \@secondoftwo
 \fi
}%
\providecommand \natexlab [1]{#1}%
\providecommand \enquote  [1]{``#1''}%
\providecommand \bibnamefont  [1]{#1}%
\providecommand \bibfnamefont [1]{#1}%
\providecommand \citenamefont [1]{#1}%
\providecommand \href@noop [0]{\@secondoftwo}%
\providecommand \href [0]{\begingroup \@sanitize@url \@href}%
\providecommand \@href[1]{\@@startlink{#1}\@@href}%
\providecommand \@@href[1]{\endgroup#1\@@endlink}%
\providecommand \@sanitize@url [0]{\catcode `\\12\catcode `\$12\catcode `\&12\catcode `\#12\catcode `\^12\catcode `\_12\catcode `\%12\relax}%
\providecommand \@@startlink[1]{}%
\providecommand \@@endlink[0]{}%
\providecommand \url  [0]{\begingroup\@sanitize@url \@url }%
\providecommand \@url [1]{\endgroup\@href {#1}{\urlprefix }}%
\providecommand \urlprefix  [0]{URL }%
\providecommand \Eprint [0]{\href }%
\providecommand \doibase [0]{https://doi.org/}%
\providecommand \selectlanguage [0]{\@gobble}%
\providecommand \bibinfo  [0]{\@secondoftwo}%
\providecommand \bibfield  [0]{\@secondoftwo}%
\providecommand \translation [1]{[#1]}%
\providecommand \BibitemOpen [0]{}%
\providecommand \bibitemStop [0]{}%
\providecommand \bibitemNoStop [0]{.\EOS\space}%
\providecommand \EOS [0]{\spacefactor3000\relax}%
\providecommand \BibitemShut  [1]{\csname bibitem#1\endcsname}%
\let\auto@bib@innerbib\@empty
\bibitem [{\citenamefont {Fall}\ \emph {et~al.}(2010)\citenamefont {Fall}, \citenamefont {Lemaitre}, \citenamefont {Bertrand}, \citenamefont {Bonn},\ and\ \citenamefont {Ovarlez}}]{fall2010shear}%
  \BibitemOpen
  \bibfield  {author} {\bibinfo {author} {\bibfnamefont {A.}~\bibnamefont {Fall}}, \bibinfo {author} {\bibfnamefont {A.}~\bibnamefont {Lemaitre}}, \bibinfo {author} {\bibfnamefont {F.}~\bibnamefont {Bertrand}}, \bibinfo {author} {\bibfnamefont {D.}~\bibnamefont {Bonn}},\ and\ \bibinfo {author} {\bibfnamefont {G.}~\bibnamefont {Ovarlez}},\ }\href@noop {} {\bibfield  {journal} {\bibinfo  {journal} {Physical review letters}\ }\textbf {\bibinfo {volume} {105}} (\bibinfo {year} {2010})}\BibitemShut {NoStop}%
\bibitem [{\citenamefont {Madraki}\ \emph {et~al.}(2020)\citenamefont {Madraki}, \citenamefont {Oakley}, \citenamefont {Nguyen~Le}, \citenamefont {Colin}, \citenamefont {Ovarlez},\ and\ \citenamefont {Hormozi}}]{madraki2020shear}%
  \BibitemOpen
  \bibfield  {author} {\bibinfo {author} {\bibfnamefont {Y.}~\bibnamefont {Madraki}}, \bibinfo {author} {\bibfnamefont {A.}~\bibnamefont {Oakley}}, \bibinfo {author} {\bibfnamefont {A.}~\bibnamefont {Nguyen~Le}}, \bibinfo {author} {\bibfnamefont {A.}~\bibnamefont {Colin}}, \bibinfo {author} {\bibfnamefont {G.}~\bibnamefont {Ovarlez}},\ and\ \bibinfo {author} {\bibfnamefont {S.}~\bibnamefont {Hormozi}},\ }\href@noop {} {\bibfield  {journal} {\bibinfo  {journal} {Journal of Rheology}\ }\textbf {\bibinfo {volume} {64}} (\bibinfo {year} {2020})}\BibitemShut {NoStop}%
\bibitem [{\citenamefont {Tapia}\ \emph {et~al.}(2022)\citenamefont {Tapia}, \citenamefont {Ichihara}, \citenamefont {Pouliquen},\ and\ \citenamefont {Guazzelli}}]{tapia2022viscous}%
  \BibitemOpen
  \bibfield  {author} {\bibinfo {author} {\bibfnamefont {F.}~\bibnamefont {Tapia}}, \bibinfo {author} {\bibfnamefont {M.}~\bibnamefont {Ichihara}}, \bibinfo {author} {\bibfnamefont {O.}~\bibnamefont {Pouliquen}},\ and\ \bibinfo {author} {\bibfnamefont {{\'E}.}~\bibnamefont {Guazzelli}},\ }\href@noop {} {\bibfield  {journal} {\bibinfo  {journal} {Physical Review Letters}\ }\textbf {\bibinfo {volume} {129}} (\bibinfo {year} {2022})}\BibitemShut {NoStop}%
\bibitem [{\citenamefont {Le}\ \emph {et~al.}(2023)\citenamefont {Le}, \citenamefont {Izzet}, \citenamefont {Ovarlez},\ and\ \citenamefont {Colin}}]{le2023solvents}%
  \BibitemOpen
  \bibfield  {author} {\bibinfo {author} {\bibfnamefont {A.~V.~N.}\ \bibnamefont {Le}}, \bibinfo {author} {\bibfnamefont {A.}~\bibnamefont {Izzet}}, \bibinfo {author} {\bibfnamefont {G.}~\bibnamefont {Ovarlez}},\ and\ \bibinfo {author} {\bibfnamefont {A.}~\bibnamefont {Colin}},\ }\href@noop {} {\bibfield  {journal} {\bibinfo  {journal} {Journal of Colloid and Interface Science}\ }\textbf {\bibinfo {volume} {629}} (\bibinfo {year} {2023})}\BibitemShut {NoStop}%
\bibitem [{\citenamefont {Trulsson}\ \emph {et~al.}(2012)\citenamefont {Trulsson}, \citenamefont {Andreotti},\ and\ \citenamefont {Claudin}}]{trulsson2012transition}%
  \BibitemOpen
  \bibfield  {author} {\bibinfo {author} {\bibfnamefont {M.}~\bibnamefont {Trulsson}}, \bibinfo {author} {\bibfnamefont {B.}~\bibnamefont {Andreotti}},\ and\ \bibinfo {author} {\bibfnamefont {P.}~\bibnamefont {Claudin}},\ }\href@noop {} {\bibfield  {journal} {\bibinfo  {journal} {Physical review letters}\ }\textbf {\bibinfo {volume} {109}} (\bibinfo {year} {2012})}\BibitemShut {NoStop}%
\bibitem [{\citenamefont {Ness}\ and\ \citenamefont {Sun}(2016)}]{ness2016shear}%
  \BibitemOpen
  \bibfield  {author} {\bibinfo {author} {\bibfnamefont {C.}~\bibnamefont {Ness}}\ and\ \bibinfo {author} {\bibfnamefont {J.}~\bibnamefont {Sun}},\ }\href@noop {} {\bibfield  {journal} {\bibinfo  {journal} {Soft matter}\ }\textbf {\bibinfo {volume} {12}} (\bibinfo {year} {2016})}\BibitemShut {NoStop}%
\bibitem [{\citenamefont {Vo}\ \emph {et~al.}(2020)\citenamefont {Vo}, \citenamefont {Nezamabadi}, \citenamefont {Mutabaruka}, \citenamefont {Delenne},\ and\ \citenamefont {Radjai}}]{vo2020additive}%
  \BibitemOpen
  \bibfield  {author} {\bibinfo {author} {\bibfnamefont {T.~T.}\ \bibnamefont {Vo}}, \bibinfo {author} {\bibfnamefont {S.}~\bibnamefont {Nezamabadi}}, \bibinfo {author} {\bibfnamefont {P.}~\bibnamefont {Mutabaruka}}, \bibinfo {author} {\bibfnamefont {J.-Y.}\ \bibnamefont {Delenne}},\ and\ \bibinfo {author} {\bibfnamefont {F.}~\bibnamefont {Radjai}},\ }\href@noop {} {\bibfield  {journal} {\bibinfo  {journal} {Nature communications}\ }\textbf {\bibinfo {volume} {11}} (\bibinfo {year} {2020})}\BibitemShut {NoStop}%
\bibitem [{\citenamefont {DeGiuli}\ \emph {et~al.}(2015)\citenamefont {DeGiuli}, \citenamefont {D{\"u}ring}, \citenamefont {Lerner},\ and\ \citenamefont {Wyart}}]{degiuli2015unified}%
  \BibitemOpen
  \bibfield  {author} {\bibinfo {author} {\bibfnamefont {E.}~\bibnamefont {DeGiuli}}, \bibinfo {author} {\bibfnamefont {G.}~\bibnamefont {D{\"u}ring}}, \bibinfo {author} {\bibfnamefont {E.}~\bibnamefont {Lerner}},\ and\ \bibinfo {author} {\bibfnamefont {M.}~\bibnamefont {Wyart}},\ }\href@noop {} {\bibfield  {journal} {\bibinfo  {journal} {Physical Review E}\ }\textbf {\bibinfo {volume} {91}} (\bibinfo {year} {2015})}\BibitemShut {NoStop}%
\bibitem [{\citenamefont {Gallier}\ \emph {et~al.}(2014)\citenamefont {Gallier}, \citenamefont {Lemaire}, \citenamefont {Peters},\ and\ \citenamefont {Lobry}}]{gallier2014rheology}%
  \BibitemOpen
  \bibfield  {author} {\bibinfo {author} {\bibfnamefont {S.}~\bibnamefont {Gallier}}, \bibinfo {author} {\bibfnamefont {E.}~\bibnamefont {Lemaire}}, \bibinfo {author} {\bibfnamefont {F.}~\bibnamefont {Peters}},\ and\ \bibinfo {author} {\bibfnamefont {L.}~\bibnamefont {Lobry}},\ }\href@noop {} {\bibfield  {journal} {\bibinfo  {journal} {Journal of Fluid Mechanics}\ }\textbf {\bibinfo {volume} {757}} (\bibinfo {year} {2014})}\BibitemShut {NoStop}%
\bibitem [{\citenamefont {Goyal}\ \emph {et~al.}(2024)\citenamefont {Goyal}, \citenamefont {Martys},\ and\ \citenamefont {Del~Gado}}]{goyal2024flow}%
  \BibitemOpen
  \bibfield  {author} {\bibinfo {author} {\bibfnamefont {A.}~\bibnamefont {Goyal}}, \bibinfo {author} {\bibfnamefont {N.~S.}\ \bibnamefont {Martys}},\ and\ \bibinfo {author} {\bibfnamefont {E.}~\bibnamefont {Del~Gado}},\ }\href@noop {} {\bibfield  {journal} {\bibinfo  {journal} {Journal of Rheology}\ }\textbf {\bibinfo {volume} {68}} (\bibinfo {year} {2024})}\BibitemShut {NoStop}%
\bibitem [{\citenamefont {Olsson}\ and\ \citenamefont {Teitel}(2007)}]{olsson2007critical}%
  \BibitemOpen
  \bibfield  {author} {\bibinfo {author} {\bibfnamefont {P.}~\bibnamefont {Olsson}}\ and\ \bibinfo {author} {\bibfnamefont {S.}~\bibnamefont {Teitel}},\ }\href@noop {} {\bibfield  {journal} {\bibinfo  {journal} {Physical review letters}\ }\textbf {\bibinfo {volume} {99}} (\bibinfo {year} {2007})}\BibitemShut {NoStop}%
\bibitem [{\citenamefont {Ness}\ \emph {et~al.}(2022)\citenamefont {Ness}, \citenamefont {Seto},\ and\ \citenamefont {Mari}}]{ness2022physics}%
  \BibitemOpen
  \bibfield  {author} {\bibinfo {author} {\bibfnamefont {C.}~\bibnamefont {Ness}}, \bibinfo {author} {\bibfnamefont {R.}~\bibnamefont {Seto}},\ and\ \bibinfo {author} {\bibfnamefont {R.}~\bibnamefont {Mari}},\ }\href@noop {} {\bibfield  {journal} {\bibinfo  {journal} {Annual Review of Condensed Matter Physics}\ }\textbf {\bibinfo {volume} {13}} (\bibinfo {year} {2022})}\BibitemShut {NoStop}%
\end{thebibliography}%

\end{document}